\documentclass[prb,twocolumn]{revtex4}

\usepackage{amsmath}

\usepackage{graphicx}

\newcommand{\rv}{{\bf r}}
\newcommand{\jv}{{\bf j}}
\newcommand{\kv}{{\bf k}}
\newcommand{\nv}{{\bf n}}
\newcommand{\pv}{{\bf p}}
\newcommand{\ev}{{\bf e}} % change macro if don't like my change

\newcommand{\Ev}{{\bf E}}

\begin{document}

\title{Reciprocal transmittances and reflectances: An elementary proof}

\author{Masanobu Iwanaga}
\email{M.Iwanaga@osa.org}
\affiliation{Department of Physics, Graduate School of Science, 
Tohoku University, Sendai 980-8578, Japan}
\author{A. S. Vengurlekar}
\affiliation{Tata Institute for Fundamental Research, Colaba, Mumbai 
400005, India}
\author{Takafumi Hatano}
\author{Teruya Ishihara}
\altaffiliation[Also at ]{Frontier Research Systems, 
RIKEN, Wako 351-0198, Japan.}
\affiliation{Department of Physics, Graduate School of Science, 
Tohoku University, Sendai 980-8578, Japan}

\date{\today}

\begin{abstract}
We present an elementary proof concerning reciprocal transmittances and 
reflectances. The proof is direct, simple, 
and valid for the diverse objects that can be absorptive and 
induce diffraction and scattering, 
as long as the objects respond linearly and locally to electromagnetic 
waves. The proof enables students who understand the basics of classical 
electromagnetics to grasp the physical basis of reciprocal optical responses. 
In addition, we show an example to demonstrate reciprocal 
response numerically and experimentally. 
\end{abstract}

\maketitle

\section{Introduction\label{intro}}
\begin{figure}[h!]
\begin{center}
\includegraphics[width=7.5cm,clip]{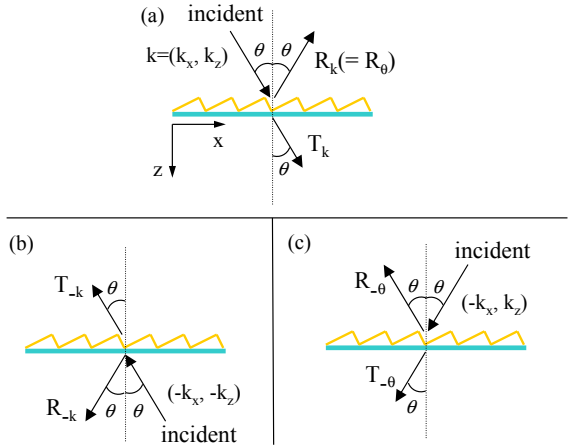}
\caption{\label{fig1}Reciprocal configurations. (a) and (b) show 
reciprocal configurations for transmission. $T_{\rm k}$ in (a) denotes 
transmittance for incident wavenumber vector $\kv$. 
$T_{\rm -k}$ in (b) is defined similarly. 
The reciprocal relation is $T_{\rm k} = T_{\rm -k}$. 
(a) and (c) are reciprocal for reflection. $R_{\theta}$ in (a) is reflectance 
for incident wavenumber vector $(k_x,k_z)$ and $R_{-\theta}$ in (c) for 
$(-k_x,k_z)$. The reciprocal relation is $R_{\theta} = R_{-\theta}$.}
\end{center}
\end{figure}

Reciprocity, which was first found by Lorentz at the end of 19th century, 
has a long history\cite{Potton} and has been derived in several formalisms. 
There are two typical reciprocal configurations in optical responses 
as shown in Fig.~\ref{fig1}. The configurations in Figs.~\ref{fig1}(a) 
and \ref{fig1}(b) are transmission reciprocal 
and those in Figs.~\ref{fig1}(a) 
and \ref{fig1}(c) are reflection reciprocal. As shown in Fig.~\ref{fig1}, 
we denote transmittance by $T$ and reflectance by $R$; the suffice k and 
$\theta$ stand for incident wavenumber vector and angle, respectively. 
The reciprocal configurations are obtained by symmetry operations 
on the incident light of the wavenumber vector: ($k_x,\, k_z) \to 
(-k_x, -k_z$) or ($-k_x,\, k_z$). 
Reciprocity on transmission means that $T_{\rm k} = T_{\rm -k}$, 
and that on reflection is expressed as $R_{\theta} = R_{-\theta}$, which is not intuitively obvious and is frequently surprising to students. 

The most general proof 
was published by Petit in 1980,\cite{Petit} where reciprocal reflection 
as shown in Fig.~\ref{fig1} is derived 
for asymmetric gratings such as an echelette grating. 
On the basis of the reciprocal relation for
the solutions of the Helmholtz equation, 
the proof showed that reciprocal reflection holds for 
periodic objects irrespective of absorption. It seems difficult 
to apply the proof to transmission because it would be 
necessary to construct solutions of Maxwell equations that satisfy 
the boundary conditions at the interfaces of the incident, grating, and 
transmitted layers. 
The history of the literature on reciprocal optical responses 
has been reviewed in Ref~\onlinecite{Potton}. 

Since the 1950s, scattering problems regarding light, elementary particles, 
and so on have been addressed by using scattering matrix (S-matrix). 
In the studies employing the S-matrix, it is assumed 
that there is no absorption by the object. The assumption leads to the 
unitarity of the S-matrix and makes it possible to prove reciprocity. 
The reciprocal reflection of lossless objects 
was verified in this formalism.\cite{Gippius}

In this paper we present a simple, direct, and general derivation 
of the reciprocal optical responses for transmission and reflection 
relying only on classical electrodynamics. 
We start from the reciprocal theorem described in Sec.~\ref{thm} and derive 
the equation for zeroth order transmission and reflection coefficients 
in Sec.~\ref{proof}. The equation is essential to the reciprocity. 
A numerical and experimental example of reciprocity is presented in 
Sec.~\ref{example}. 
The limitation and break down of reciprocal optical responses 
are also discussed. 

\section{Reciprocal Theorem\label{thm}}
The reciprocal theorem has been proved in various fields, such as statistical 
mechanics, quantum mechanics, and electromagnetism.\cite{Landau} 
Here we introduce the theorem for electromagnetism. 

\begin{figure}[h!]
\begin{center}
\includegraphics[width=7cm,clip]{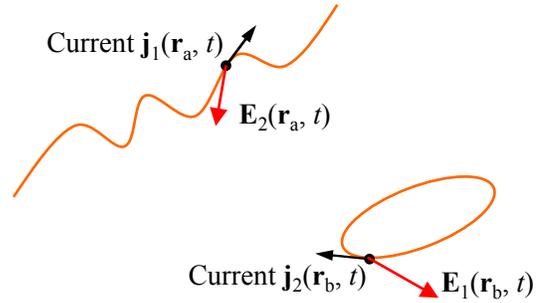}
\caption{\label{fig2}Schematic drawing of two 
currents $\jv_i$ and the electric fields $\Ev_i$ induced by 
$\jv_i$ ($i=1,2$). The curves denote the position where the currents exist.}
\end{center}
\end{figure}

When two currents exist as in Fig.~\ref{fig2} and the induced 
electromagnetic (EM) waves travel in linear and locally responding media 
in which 
$D_i(\rv) = \sum_j \varepsilon_{ij}E_j(\rv)$ and 
$B_i(\rv) = \sum_j \mu_{ij}H_j(\rv)$, then 
\begin{equation}
\int \jv_1(\rv)\cdot \Ev_2(\rv) d\rv = 
\!\int \jv_2(\rv)\cdot \Ev_1(\rv) d\rv.\label{reci}
\end{equation}
Equation~\eqref{reci} is the reciprocal theorem in electromagnetism. 
The proof shown in 
Ref.~\onlinecite{Landau} exploits plane waves and is straightforward. 
Equation~(\ref{reci}) is valid even for media with losses. 
The integrands take non-zero values at the position $\rv$ 
where currents exist, that is, $\jv_i(\rv)\neq{\bf 0}$. 
The theorem indicates the reciprocity between 
the two current sources $\jv_i$ ($i = 1,2$) and the induced EM waves $\Ev_i$ 
which are observed at the position of the other source $\jv_k$ 
($k \neq i$). 

\section{Reciprocal Optical Responses\label{proof}}
In this section, we apply the reciprocal theorem to optical responses 
in both transmission and reflection configurations. 
First, we define the notation used in the calculations of the integrals in 
Eq.~(\ref{reci}).
An electric dipole oscillating at the frequency $\omega$ 
emits dipole radiation, which is detected in the far field. When 
a small dipole $\pv$ along the $z$ axis is located at the origin, 
it is written as 
$\pv(t)=p(t)\ev_z$ and $p(t) = p_0 e^{i\omega t}$, where $\ev_z$ denotes 
the unit vector along the $z$ axis and $p_0$ the magnitude of the dipole. 
The dipole in vacuum emits radiation, which in the far field is 
\begin{subequations}
\begin{align}
\Ev(\rv,t) &= \frac{1}{4\pi\varepsilon_0}
\frac{\ddot{p}(t')}{c^2 r}\sin\theta \cdot\ev_{\theta} \\
&= \frac{-1}{4\pi\varepsilon_0}\frac{p_0\,\omega^2}{c^2 r}
e^{i\omega t'}\sin\theta \cdot\ev_{\theta} , \label{dipole}
\end{align}
\end{subequations}
where polar coordinates ($r$, $\theta$, $\phi$) are used, a unit vector is 
given by 
$\ev_{\theta} = (\cos\theta\cos\phi,\cos\theta\sin\phi,-\sin\theta)$, 
and $t'= t-r/c$. Because the dipole $\pv$ is defined by 
$\pv(\rv,t) = \!\int \rv\rho(\rv,t) d\rv$ 
and conservation of charge density is given by 
$\nabla\cdot\jv + \partial\rho/\partial t = 0$, we obtain 
the current $\jv$ associated with the dipole $\pv$: 
\begin{equation}
\jv(\rv,t) = \dot{p}(t)\delta(\rv)\ev_z. \label{j}
\end{equation}

\begin{figure}[h!]
\begin{center}
\includegraphics[width=6.5cm,clip]{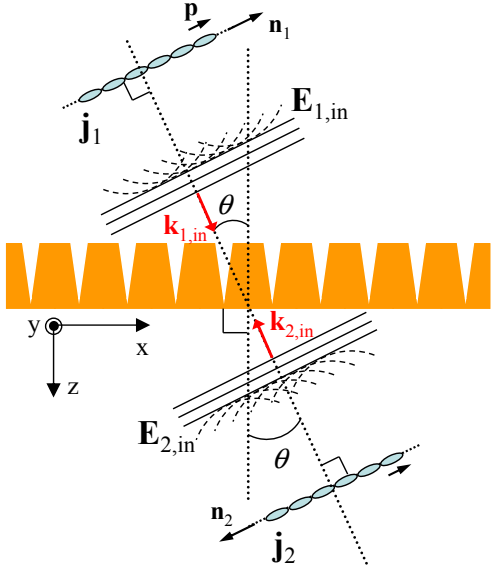}
\caption{\label{fig3}Schematic drawing of reciprocal configuration 
for transmission. The object has an arbitrary periodic structure, 
which is asymmetric along the $z$ axis. Currents $\jv_i$ induce 
electric fields $\Ev_{i,{\rm in}}$ ($i=1,2$).}
\end{center}
\end{figure}

Consider two arrays of $N$ dipoles (long but finite) 
in the $xz$ plane as shown in Fig.~\ref{fig3}. 
The two arrays have the same length, and the directions are specified 
by normalized vectors $\nv_i$ ($i=1,2$) and 
$\nv_1 \parallel \nv_2$. 
In this case, the current is $\jv_i\parallel \nv_i$. 
If the dipoles coherently oscillate with the same phase, 
then the emitted electric fields are superimposed and form a wave front 
at a position far from the array in the $xz$ plane as drawn 
in Fig.~\ref{fig3}. 
The electric field vector of the wave front, $\Ev_{i,{\rm in}}$, 
satisfies $\Ev_{i,{\rm in}}\parallel \nv_i$ 
and travels with wavenumber vector $\kv_{i,{\rm in}}$. 
Thus, if we place the dipole arrays far enough from the object, 
the induced EM waves become slowly decaying incident plane 
waves in the $xz$ plane to a good approximation. 
The arrays of dipoles have to be long enough to form the plane wave. 

For the transmission configuration, we calculate 
$\int \jv_i\cdot \Ev_k\,d\rv$ ($i,k=1,2$ and $i \neq k$). 
Figure~\ref{fig3} shows a typical transmission configuration, which includes 
an arbitrary periodic object asymmetric along the $z$ axis. The relation 
between the current $\jv_i$, the direction $\nv_i$ of the dipole, 
and the wavenumber vector $\kv_{i,{\rm in}}$ of the wave front is 
summarized as $\jv_i \parallel \nv_i$ and 
$\nv_i \perp \kv_{i,{\rm in}}$. 
It is convenient to expand the electric field into a Fourier series 
for the calculation of periodic sources: 
\begin{equation}
\Ev(\rv) = \sum_m \Ev^{(m)}
\exp(i\kv_m\cdot \rv), \label{E_expand}
\end{equation}
where $\Ev^{(m)}$ is the Fourier coefficient of $\Ev(\rv)$, 
$\kv_m = (k_{x,m},0,k_{z,m}) = (k_{{\rm in},x}+2\pi m/d_x, 0, k_{z,m})$ 
($m=0,\pm 1,\pm 2, \cdots$), and $d_x$ is the periodicity of the object 
along the $x$ axis (see Fig.~\ref{fig3}). 
The $z$ component is expressed in homogeneous media in vacuum as 
$k_{z,m} = \pm\sqrt{\kv_{\rm in}^2 - k_{x,m}^2}$, where 
the signs correspond to the directions along the $z$ axis. 

When the dipole array is composed of sufficiently small and 
numerous dipoles, the integration can be calculated to good accuracy as
\begin{subequations}
\begin{align}
\int \jv_1(\rv) \cdot \Ev_2(\rv) d\rv 
&= \!\int i\omega p_0 \nv_1 \cdot \sum_m \Ev_2^{(m)}
\exp(i\kv_m\cdot s \nv_1) ds \\
&= \sum_m \delta_{m,0} N(i\omega p_0 \nv_1 \cdot 
\Ev_2^{(m)}) \\
&= i\omega N p_0 E_2^{(0)}, \label{j1E2}
\end{align}
\end{subequations}
where $E_2^{(0)} = |\Ev_2^{(0)}|$. 
To ensure that the integration is proportional to $\delta_{m,0}$, 
the array of dipoles has to be longer than $L$: 
\begin{equation}
L = (\mbox{length of dipole})\cdot q,
\end{equation}
where $q$ is the least common multiple of the diffraction channels which are 
open at the frequency $\omega$. 
This condition would usually be satisfied when $\Ev_{i,{\rm in}}$ 
forms a plane wave. 

By permutating 1 and 2 in Eq.~(\ref{j1E2}), we obtain 
$\int \jv_2\cdot\Ev_1 d\rv = i\omega N p_0 E_1^{(0)}$. 
Equation~(\ref{j1E2}) and the reciprocal theorem in 
Eq.~(\ref{reci}) lead to the equation 
\begin{equation}
E_1^{(0)} = E_2^{(0)}. \label{E_reci}
\end{equation}
Each electric vector $E_i^{(0)}$ ($i=1,2$) is observed at the position $\rv$ 
where there is another current $\jv_k(\rv)$ ($k\neq i$). 
The integral in Eq.~(\ref{reci}) is reduced to 
Eq.~(\ref{j1E2}) which is expressed only by the zeroth components of the 
transmitted electric field. The reciprocity is thus independent of higher 
order harmonics, 
which are responsible for the modulated EM fields in structured objects. 
When there is no periodic object in Fig.~\ref{fig3}, 
a similar relation holds: 
\begin{equation}
E_1^{\rm no,(0)} = E_2^{\rm no,(0)}. \label{E0_reci}
\end{equation}
The transmittance $T_i$ is given by 
\begin{equation}
T_i = \left|\frac{E_i^{(0)}}{E_i^{\rm no,(0)}}\right|^2.\label{T_reci}
\end{equation}
From Eqs.~(\ref{E_reci})--(\ref{T_reci}), 
we finally reach the reciprocal relation $T_1 = T_2$. 

The feature of the proof that $T_1 = T_2$ is independent of the detailed 
evaluation of $E_i^{(0)}$ and therefore makes the proof simple and general. 
The proof can be extended to two-dimensional periodic structure by 
replacing the one-dimensional periodic structure in Fig.~\ref{fig3} 
by two-dimensional one. 
Although we have considered periodic objects, the proof can also 
be extended to non-periodic objects. To do this extension, Eq.~(\ref{E_expand}) has to be expressed in the general form 
$\Ev(\rv) = \int \Ev(\kv)\exp(i\kv\cdot\rv)d\kv$, 
and a more detailed calculation for 
$\int\jv_i\cdot\Ev_kd\rv$ is required. 
Reciprocity for transmission thus holds 
irrespective of absorption, diffraction, and scattering by objects. 

In Fig.~\ref{fig3} the induced electric fields $\Ev_i$ are 
polarized in the $xz$ plane. The polarization is called TM polarization 
in the terminology of waveguide theory and is also often called $p$ 
polarization. 
For TE polarization (which is often called $s$ polarization) 
for which $\Ev_i$ has a polarization 
parallel to the $y$ axis, the proof is similar to what we have described 
except that the dipoles are aligned along the $y$ axis. 

\begin{figure}[h!]
\begin{center}
\includegraphics[width=7cm,clip]{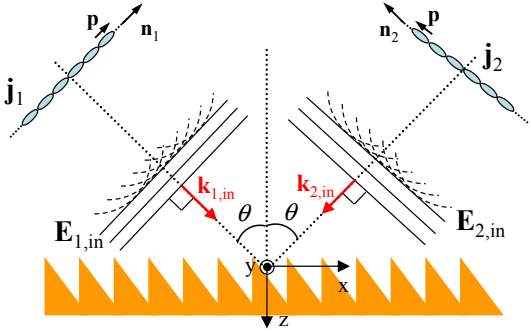}
\caption{\label{fig4}Schematic configuration for reciprocal reflection. 
The object has an arbitrary periodic structure, 
which consists of asymmetric unit cells. 
The currents $\jv_i$ yield electric fields $\Ev_{i,{\rm in}}$ 
($i=1,2$).}
\end{center}
\end{figure}

Reciprocal reflection is also shown in a similar way. 
The configuration is depicted in Fig.~\ref{fig4}. The two sources have to be 
located to satisfy the mirror symmetry about the $z$ axis. 
The calculation of $\int\jv_i\cdot\Ev_kd\rv$ leads 
to the reciprocal relation for reflectance $R_1 = R_2$. 
Note that $E_i^{{\rm no},(0)}$ in Eq.~(\ref{E0_reci}) has to be evaluated 
by replacing the periodic object by a perfect mirror. 

\section{Numerical and Experimental Confirmation\label{example}}
\begin{figure}[h!]
\begin{center}
\includegraphics[width=7cm,clip]{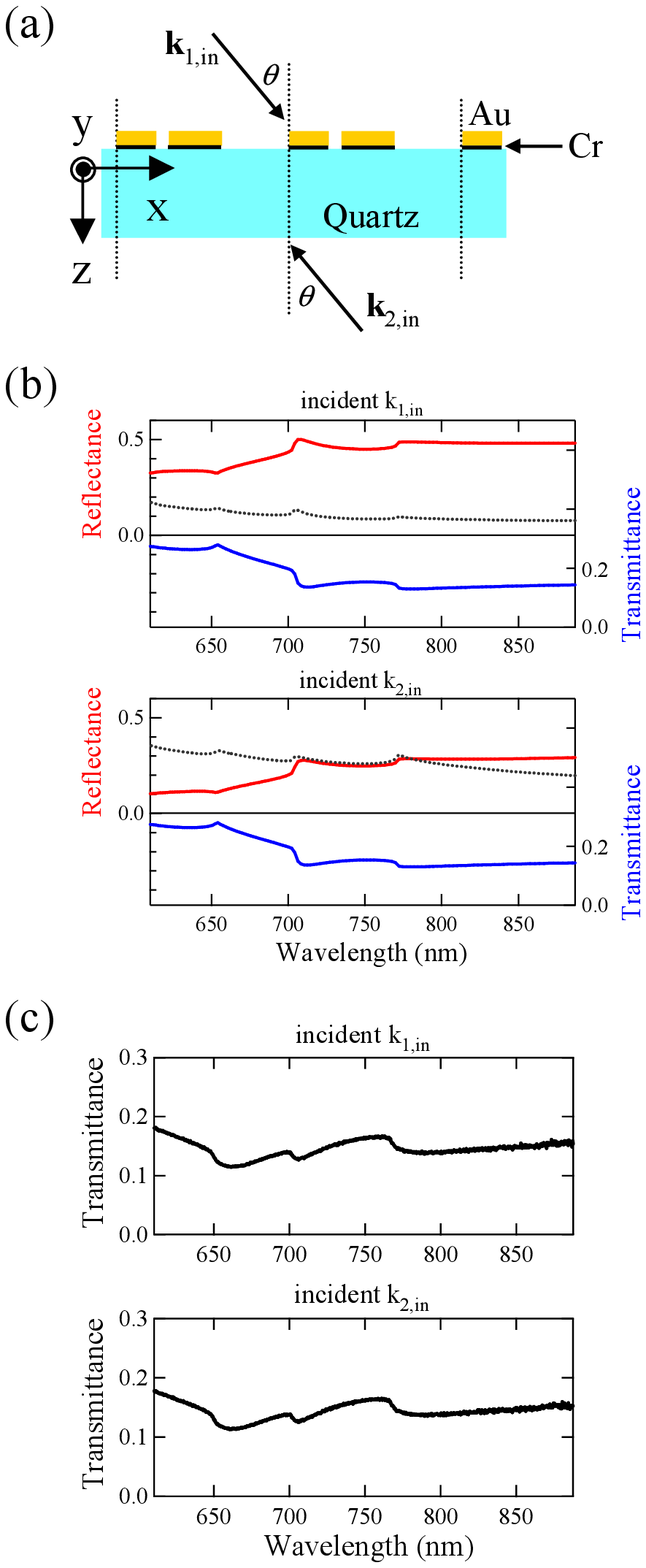}
\caption{\label{fig5}(a) Schematic drawing of metallic grating profile 
modeled from AFM images. The periodicity is 1200\,nm. 
The dotted lines show the unit cells in which the ratio is Au:air:Au:air = 
3:1:4:5. The thickness of Au, Cr, and the quartz substrate is 40\,nm, 
5\,nm, and 1\,mm, respectively. (b) Numerically calculated 
spectra for 10$^{\circ}$ incidence of $\kv_{1,{\rm in}}$ 
(upper panel) and $\kv_{2,{\rm in}}$ (lower panel) of TM polarization. 
In both panels the reflectance (upper solid line) and absorption (dotted 
line) are plotted using the left axis, while the transmittance (lower solid 
line) uses the right axis. 
(c) Measured transmittance spectra, corresponding to the transmittance 
spectra in (b).}
\end{center}
\end{figure}

An example of reciprocal optical response is shown here. 
Figure \ref{fig5}(a) displays the structure of the sample and 
reciprocal transmission configuration. The sample consists of 
periodic grooves etched in metallic films of Au and Cr on a quartz substrate. 
The periodicity is 1200\,nm, as indicated by the dotted lines in 
Fig.~\ref{fig5}(a). 
The unit cell has the structure of Au:air:Au:air = 3:1:4:5. 
The thickness of Au, Cr, and quartz is 40\,nm, 
5\,nm, and 1\,mm, respectively. 
The structure is obviously asymmetric about the $z$ axis. 
The profile was modeled from an AFM image of the fabricated sample. 

Figure~\ref{fig5}(b) shows our numerical results. 
The incident light has $\theta = 10^{\circ}$ and TM polarization (the electric 
vector is in the $xz$ plane). 
The numerical calculation was done with an 
improved S-matrix method\cite{Tikh, Li1} The permittivities of gold and 
chromium were taken from Refs.~\onlinecite{Johnson} and 
\onlinecite{Johnson2}; the permittivity of quartz is well known to be 2.13. 
In the numerical calculation, the incident light is taken to be a plane wave, 
and harmonics up to $n=\pm 75$ in Eq.~(\ref{E_expand}) were used, which
is enough to obtain accurate optical responses. 
The result indicates 
that transmission spectra (lower solid line) are numerically 
the same in the reciprocal configurations, while reflection (upper solid 
line) and absorption (dotted line) spectra show a definite difference. The 
absorption is plotted along the left axis. The difference 
implies that surface excitations are different on each side and absorb 
different numbers of photons.
Nonetheless, the transmission spectra are the same for incident 
wavenumber vectors $\kv_{1,{\rm in}}$ and $\kv_{2,{\rm in}}$. 

Experimental transmission spectra are shown in Fig.~\ref{fig5}(c) and 
are consistent within experimental error. 
Reciprocity is thus confirmed both numerically and experimentally. 
There have a few experiments on reciprocal transmission (see references 
in Ref.~\onlinecite{Potton}). In comparison with these results, Fig.~\ref{fig5}(c) shows the excellent agreement of reciprocal transmission and is 
the best available experimental evidence supporting reciprocity. 

We note that transmission spectra in Figs.~\ref{fig5}(b) and \ref{fig5}(c) 
agree quantitatively above 700\,nm. On the other hand, they show a qualitative 
discrepancy below 700\,nm. The result could come from the difference between 
the modeled profile in Fig.~\ref{fig5}(a) and the actual profile of the 
sample. 
The dip at 660\,nm stems from a surface plasmon at the metal-air 
interface, so that the measured transmission spectra would be affected 
significantly by the surface roughness and the deviation from 
the modeled structure.

\section{Remarks and summary\label{summary}}
As described in Sec.~\ref{thm}, the reciprocal theorem assumes that 
all media are linear and show local response. Logically, it can happen that 
the reciprocal optical responses do not hold for nonlinear or 
nonlocally responding media. 

Reference~\onlinecite{non-recipro} discusses an explicit difference of the 
transmittance for a reciprocal 
configuration in a nonlinear optical crystal of KNbO$_3$:Mn. 
The values of the transmittance deviate by a few tens of percent in the 
reciprocal configuration. 
The crystal has a second-order response such that 
$D_i(\rv) = \sum_j \varepsilon_{ij}E_j(\rv) + \sum_{j,k}\varepsilon_{ijk}
E_j(\rv)E_k(\rv)$.
The break down of reciprocity comes from the nonlinearity. 

Does reciprocity also break down in nonlocal media? 
In nonlocal media the induction \textbf{D} is given by 
$\textbf{D}(\rv) = \!\int\!\varepsilon(\rv,\rv')\Ev(\rv')d\rv'$. Although 
a general proof for this case has not been reported to our knowledge, 
it has been shown that reciprocity holds in a particular stratified structure 
composed of nonlocal media.\cite{H.Ishihara} 

In summary, we have presented an elementary and heuristic proof of the 
reciprocal optical responses for transmittance and reflectance. 
When the reciprocal theorem in Eq.~(\ref{reci}) holds, the reciprocal 
relations come from geometrical configurations of light sources and 
observation points, and are independent of the details of the objects. 
Transmission reciprocity has been confirmed both numerically
and experimentally. 

\begin{acknowledgments}
We thank S.\ G.\ Tikhodeev for discussions. One of us (M.\ I.) 
acknowledges the Research Foundation for Opto-Science and Technology for 
financial support, and the Information Synergy Center, Tohoku University for 
their support of the numerical calculations. 
\end{acknowledgments}

\newpage

\section*{Figure Captions}

\begin{thebibliography}{20}
\bibitem{Potton}R. J. Potton,``Reciprocity in optics,'' 
Rep. Prog. Phys. {\bf 67}, 717--754 (2004).

\bibitem{Petit}R. Petit, ``A tutorial introduction,'' 
in \textit{Electromagnetic Theory of Gratings}, 
edited by R. Petit (Springer, Berlin, 1980), p. 1.

\bibitem{Gippius}N. A. Gippius, S. G. Tikhodeev, and 
T. Ishihara, ``Optical properties of photonic crystal slabs with an 
asymmetric unit cell,'' Phys. Rev. B {\bf 72}, 045138-1--7 (2005).

\bibitem{Landau}L. D. Landau, E. M. Lifshitz, and L. P. Pitaevskii, 
\textit{Electrodynamics of Continuous Media} (Pergamon Press, 
NY, 1984), 2nd ed.

\bibitem{Jackson}J. D. Jackson, \textit{Classical Electrodynamics} 
(John Wiley \& Sons, NJ, 1999), 3rd ed. 

\bibitem{Tikh}S. G. Tikhodeev, A. L. Yablinskii, E. A. Muljarov, 
N. A. Gippius, and T. Ishihara, ``Quasiguided modes and optical properties 
of photonic crystal slabs,'' 
Phys. Rev. B {\bf 66}, 045102-1--17 (2002).

\bibitem{Li1}L. Li, ``Use of Fourier series in the analysis of discontinuous 
periodic structures,''
J. Opt. Soc. Am. A, {\bf 13}, 1870--1876 (1996). 

\bibitem{Johnson}P. B. Johnson and R. W. Christy, ``Optical constants of 
the noble metals,'' Phys. Rev. B {\bf 6}, 4370--4379 (1972).

\bibitem{Johnson2}P. B. Johnson and R. W. Christy, ``Optical constants of 
transition metals: Ti, V, Cr, Mn, Fe, Co, Ni, and Pd,'' 
Phys. Rev. B {\bf 9}, 5056--5070 (1974).

\bibitem{non-recipro}M. Z. Zha and P. G\"unter,
``Nonreciprocal optical transmission through photorefractive
KNbO$_3$:Mn,'' Opt. Lett. {\bf 10}, 184--186 (1985).

\bibitem{H.Ishihara}H. Ishihara, ``Appearance of novel nonlinear optical 
response by control of excitonically resonant internal field,'' in 
\textit{Proceedings of 5th Symposium of Japanese 
Association for Condensed Matter Photophysics} (1994), pp. 287--281 
(in Japanese).

\end{thebibliography}
\end{document}